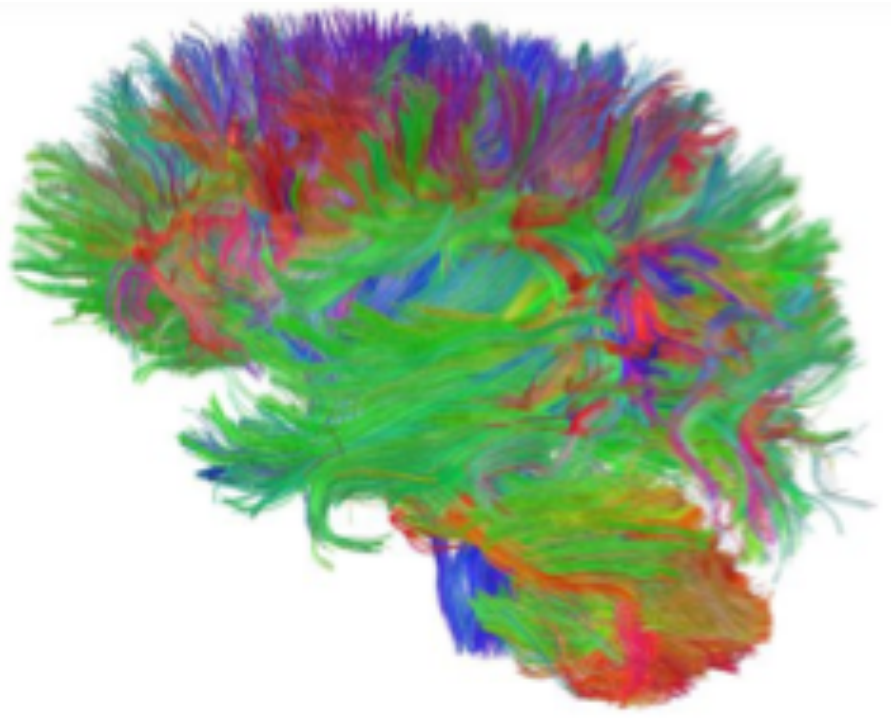

Marcus Kaiser

# Functional compensation after lesions: Predicting site and extent of recovery

# Functional compensation after lesions: Predicting site and extent of recovery

Marcus Kaiser[1,2*]

1 School of Computing Science, Newcastle University, United Kingdom
2 Riu Jin Hospital, Shanghai Jiao Tong University, China

**Correspondence:**
Dr Marcus Kaiser
School of Computing
Newcastle University
Newcastle upon Tyne, NE4 5TG
United Kingdom
m.kaiser@ieee.org

**Summary:**
In some cases the function of a lesioned area can be compensated for by another area. However, it remains unpredictable if and by which other area a lesion can be compensated. We assume that similar incoming and outgoing connections are necessary to encode the same function as the damaged region. The similarity can be measured both locally using the matching index and looking at a more global scale by non-metric multidimensional scaling (NMDS). We tested how well both measures can predict the compensating area for the loss of the visual cortex in kittens. Here, the global comparison of connectivity turns out to be a better method for predicting functional compensation. In future studies, the extent of the similarity between the lesioned and compensating regions might be a measure of the extent to which function can be successfully recovered.

## 1. Introduction

To what extent the brain can recover from lesions depends not only on the lesion itself but also on the capability of one's brain network to functionally compensate for the affected brain tissue. The concept of degeneracy describes the ability of other brain regions to perform the same function or to yield the same output as the lesioned regions (Tononi et al., 1999). A measure for degeneracy is its order (Price and Friston, 2002); that means the minimum number of regions that need to be removed before a function is lost (first-order: one region, second-order: two regions, etc.).

If a function can be retained after a lesion, even though function performance may be reduced, this is known as functional compensation. Compensation without gross rewiring of brain pathways is referred to as recovery whereas compensation at earlier developmental stages caused by unmasking of existing pathways or altered development of brain pathways is called sparing of functions (Payne and Lomber, 2001).

A necessity for the compensating area seems to be similar incoming and outgoing connections. For example, a region that compensates for the loss of the visual cortex regions should get incoming *visual* information (e.g. from the lateral geniculate body or the superior colliculus).

The cat has been discussed as a model for the study of compensatory plasticity after early vision loss. Experiments by Spear *et al.* (1980) have shown that the loss of the visual cortex (areas 17, 18 and 19) in the first weeks after birth can be compensated for by the posteromedial lateral suprasylvian area (PMLS, cf. Fig. 1). After lesion of the visual cortex an encoding of direction sensitivity and ocular dominance was found by electrophysiology in PMLS. While these functions are pre-existing to a lesser extent before the lesion, others as orientation selectivity develop *de novo* (Spear et al., 1988).

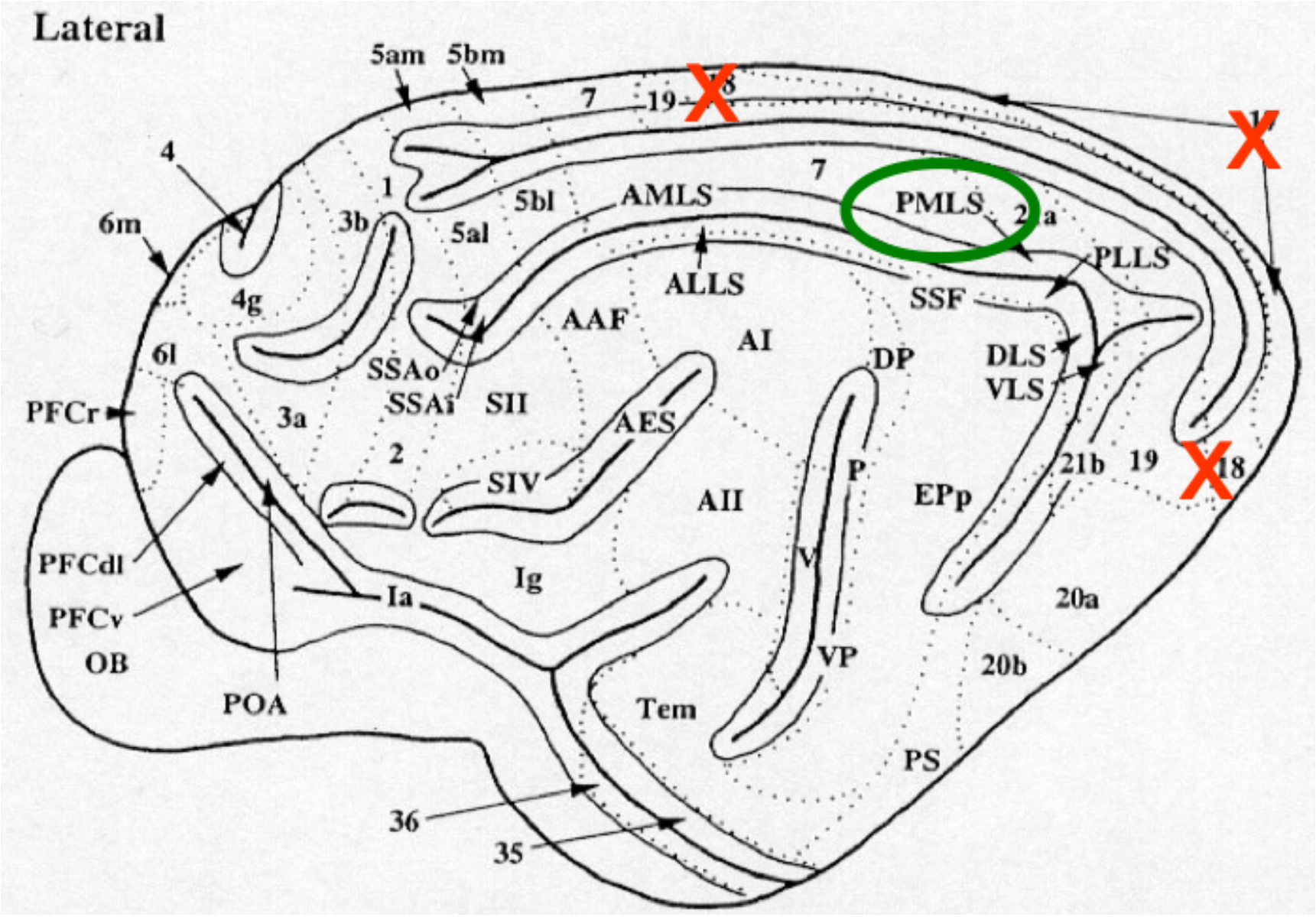

**Figure 1.** Lateral view of the cat cortex (adapted from Scannell et al., 1995). The location of the lesioned visual cortex (areas 17, 18 and 19, marked with red X signs) is shown. PMLS (green circle) is the region that exhibits functional compensation of the damaged visual areas.

The effect of lesions on functional systems has been examined before (Young et al., 2000) but here we want to discuss two ways of predicting compensating areas based on similar connectivity. First, similar areas can be found by looking at the percentage of identical direct connections calculated by the matching index or Jaccard coefficient (see Methods). Second, taking into account the connectivity relative to the whole network is used as a global measure of connection similarity. We test the ability of both measures to predict the compensation for the loss of the visual cortex in kittens.

## 2. Materials and Methods

To measure the local similarity of two areas the matching index is used (Hilgetag et al., 2000). It is defined as the number of existing identical connections divided by the number of connection pairs for which at least one connection exists (excluding direct connections between the two areas). The algorithm for the matching index was implemented in MATLAB (Release 12).

To measure the similarity based on the connectivity of the whole network non-metric multidimensional scaling (NMDS) is used. NMDS constructs a spatial representation of a set of elements on the basis of a table of 'proximities' that define the relations between the elements. We refer to the distance between regions in the three-dimensional output space as the NMDS-distance (Young et al., 1995). A connection with low NMDS-distance stands for a connection that links regions with similar connection pattern. For the calculation of the NMDS output space the ALSCAL-algorithm (Takane et al., 1977) within SPSS was used.

We applied the connection data of the cat cerebral cortex, amygdala and hippocampus (Scannell et al., 1995). The data consisted of 65 regions and 1,139 reported connections.

## 3. Results

The matching index ranks area 21a on top (matching index = 0.525). The compensating area PMLS follows shortly thereafter (matching index=0.512) on the second place. In contrast, NMDS yields PMLS on the first place with a large difference from the second place. The average NMDS-distance between PMLS and the areas 17, 18 and 19 has been 0.33 (cf. Fig. 2). This was the lowest NMDS-distance of a region to the visual cortex not only within the lateral suprasylvian area (LS) but of all remaining 62 regions excluding visual cortex. There are two regions that also have a low NMDS-distance to the visual cortex: The area 21a (average NMDS-distance: 0.5) that lies anatomically superior to PMLS and, to a lesser extent, the ventrolateral suprasylvian area (VLS, average NDMS-distance 0.56). To our knowledge there has been no research about compensation in these two areas. Also, it remains unclear if compensation in regions with high NMDS-distance to the visual cortex (e.g. posterolateral, anteromedial, anterolateral suprasylvian area) can be found.

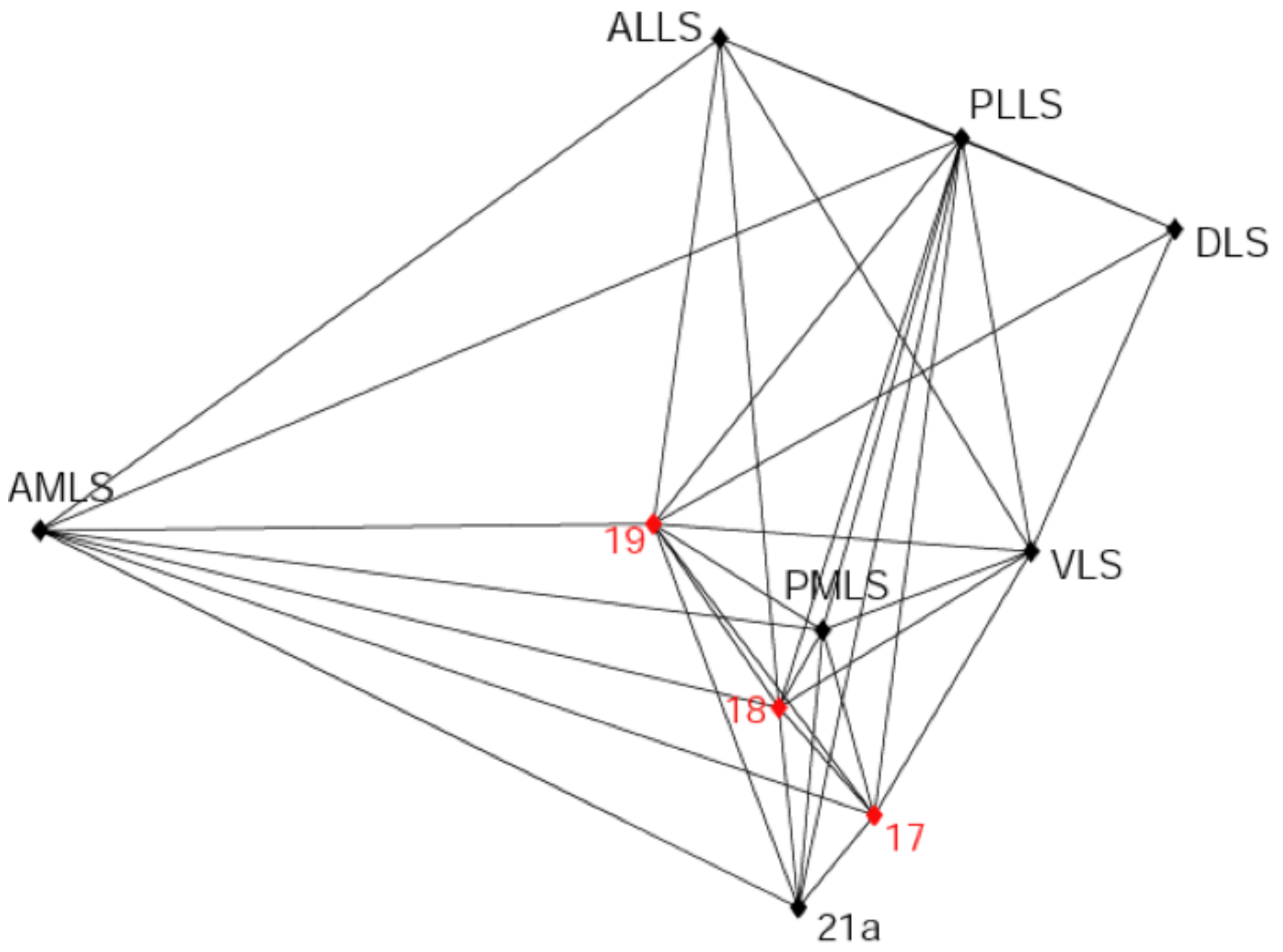

**Figure 2.** Two-dimensional projection of the location of the regions in the NMDS output space. The connections between the regions are shown. PMLS is near to the removed visual areas 17, 18 and 19 (shown in red). Also, regions VLS and 21a are near the visual cortex indicating a higher similarity of their connectivity profiles.

Therefore, comparing the direct connection pattern yields a different ranking which would predict another area for compensation. Besides predicting the correct area for compensation NDMS also has a larger difference between first and second place in the ranking. The matching index for PMLS shows that only 50% of matching connections can be sufficient for compensation during early development.

**4. Discussion**

Both local and global connectivity can be used to find similar regions for lesioned regions. However, in the case of compensating for the loss of the visual cortex in the cat only the global method (NMDS) was able to correctly predict the compensating area.

An advantage of NMDS over using the matching index is that the connection pattern of neighboring regions is also included in the calculation. This might lead to better results than comparing only the *local* efferent and afferent connections of two regions (Fig. 3).

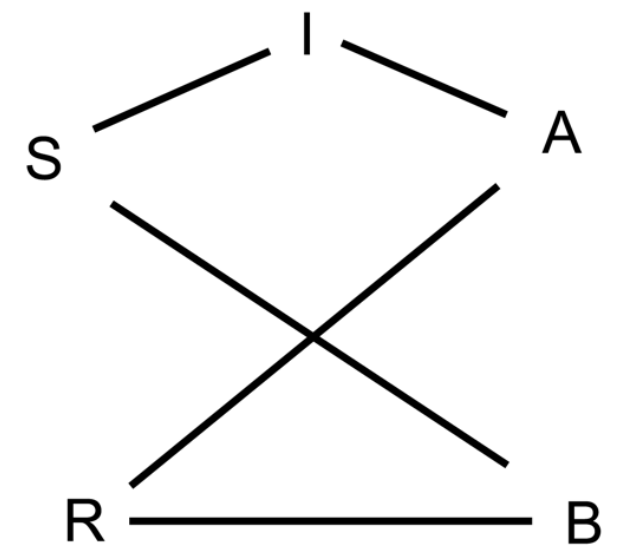

**Figure 3.** A case explaining different results for the matching index and NMDS. Areas A and B have a direct connection to R but only area B has a direct connection to area S, too. Area A though is able to get information from area S over the intermediate region I. As the matching index compares the direct connections it would underestimate the ability of area A to get information from the same regions (S and R) as area B. NMDS as referring to the global connection pattern would include that aspect.

A disadvantage concerning our data set is that each existing connection has the same value in the connectivity matrix, although some connections might be more important for a specific function than others. On the other hand, the connection pattern could be optimised in a way that all connections play an important role in the function of a region, therefore being of equal importance.

In conclusion, we demonstrated that the area in the cat cortex that compensates for the loss of the visual cortex has the lowest NMDS-distance to it. Comparing the NMDS-distance between regions could accelerate the search for regions that provide the structural correlate for functional compensation. Knowing regions that can compensate functions after lesions could be of clinical use as training and activation of such regions through brain stimulation could therefore enhance the performance of disturbed functions. In addition to identifying the region that is involved in functional compensation, the distance to the removed regions in NMDS space might be an indicator of the extent to which function can be recovered. Compensating regions with a low distance to the lesioned regions might be able to almost fully recover the initially lost function whereas a large distance, meaning that no regions with a similar connectivity pattern exist, might indicate an almost complete loss of function.

**Acknowledgement**

Marcus Kaiser was supported by Wellcome Trust (102037), Engineering and Physical Sciences Research Council (NS/A000026/1, EP/N031962/1), Medical Research Council (MR/T004347/1), and the Guangci Professorship Program of Ruijin Hospital (Shanghai Jiao Tong Univ.).